\documentstyle[12pt]{article}
\begin{document}
\def\preprint#1{\gdef\@preprint{#1}}
\def\bce{\begin{center}}
\def\ece{\end{center}}
\def\be{\begin{equation}}
\def\ee{\end{equation}}
\def\bea{\begin{eqnarray}}
\def\eea{\end{eqnarray}}
\def\thint{\int d^3\! x\,\,}
\def\ggtmn{g^{\mu\nu}}
\def\ggbmn{g_{\mu\nu}}
\def\pbm{\partial_\mu}
\def\pbn{\partial_\nu}
\def\pbr{\partial_\rho}
\def\pbs{\partial_\sigma}
\def\tbmn{T_{\mu\nu}}
\def\ggtrs{g^{\rho\sigma}}
\hskip 11cm SNUTP-96-053
\begin{center}
\def\thefootnote{\fnsymbol{footnote}}
{\bf Production of Global Monopoles through the Nucleation of $O(3)$ 
Bubbles\footnote{Talk presented at Pacific Conference on Gravitation and
Cosmology, Seoul, Feb. 1-6, 1996.}}
\vskip 1cm
\def\thefootnote{\arabic{footnote}}
Yoonbai Kim\footnote[1]{e-mail: yoonbai$@$top.phys.pusan.ac.kr},
Kei-ichi Maeda\footnote[2]{e-mail: maeda$@$cfi.waseda.ac.jp},
Nobuyuki Sakai\footnote[3]{e-mail: sakai$@$cfi.waseda.ac.jp}
\vskip 1cm
{\it ${}^{1}$Department of Physics, Pusan National University,
Pusan 609-735, Korea\\
${}^{2,3}$Department of Physics, Waseda University, Shinjuku-ku, Tokyo 169,
Japan}
\end{center}
\vskip 1cm
\noindent {\bf Abstract}
\vskip 0.7cm
First-order cosmological phase transitions are considered in the models with
an $O(3)$-symmetric scalar field, in the high temperature limit.
It is shown that a global monopole can be produced at the center of
a bubble when the bubble is nucleated.

\vskip 1cm
\noindent {\bf I. Introduction}
\vskip 0.7cm
Symmetries believed to exist in the early universe will break spontaneously as the
universe expands and cools. Such cosmological phase transitions lead to the
formation of defects - domain walls, strings, monopoles and textures - and the
dominant channel is the Kibble mechanism \cite{Kib}. The consequences of the
production of defects may be  a crisis like the over-production of magnetic
monopoles in grand unified theories, which can be resolved by inflation, while
cosmic strings may provide a viable fluctuation spectrum for galaxy formation
\cite{Rev}. Moreover, A ``new inflationary" universe can be formed at the site of
a defect itself \cite{Vil}. When the cosmological phase transition is first
order, it is achieved through the nucleation, evolution and percolation of
bubbles, and defects are generated by the collisions of bubbles. Since the
pattern of nucleated vacuum bubbles are not significantly affected by the
symmetries of the model, for a wide class of systems, the bounce solution with
one and only one negative mode of the Euclidean action can depict the vacuum
bubble solution \cite{Col}.

In this report, we review how an $O(3)$ symmetric model can support new bubble
solutions, despite the above restriction: a stable global monopole can be
simultaneously generated at their center \cite{KMS}. (For the short versions,
refer \cite{SKM} and, for the flat spacetime case, see also \cite{Kim}.)

\vskip 1cm
\noindent {\bf II. Nucleation of Monopole-Bubbles}
\vskip 0.7cm
The model of interest is an $O(3)$ symmetric scalar multiplet in a curved
spacetime. At finite temperature it is formulated using the imaginary time
method, and the Euclidean action is given by
\begin{equation}\label{eac}
S=\int^{1/T}_{0}dt_E\thint\!\sqrt{g}\;\biggl\{-\frac{1}{16\pi G}R
+\frac{1}{2}\ggtmn\pbm\phi^{a}\pbn\phi^{a}+V(\phi)\biggr\}, \end{equation}
where $\phi^{a}=\hat{\phi}^{a}\phi$ is an $O(3)$-symmetric isovector 
$(a=1,2,3)$ with $\phi=\sqrt{\phi^{a}\phi^{a}}$.
Our argument in the following does not
depend on the detailed form of the scalar potential and the existence of the
new bubble solution is guaranteed under any potential which includes one false 
vacuum and one true vacuum. Here we choose a representative $\phi^{6}$ potential
for the actual calculation,
\begin{equation}\label{pot}
V(\phi)=\frac{\lambda}{v^{2}}(\phi^{2}+\alpha v^{2})(\phi^{2}-v^{2})^{2} ~~ {\rm
with} ~~ 0<\alpha<1/2. \end{equation}
This potential is appropriate for the first order transition from a symmetric
vacuum (de Sitter spacetime with the horizon $H^{-1}\equiv(8\pi
GV(0)/3)^{-\frac{1}{2}}$) to the broken vacuum (Minkowski spacetime).

Let us consider the situation with temperature $T$, such that $1/({\rm
bubble}\;{\rm radius}) \ll T\ll M_{Pl}(=1/\sqrt{G})$, although the supermassive
scale will also raise questions about the inner structure of bubbles in the very
early universe. At sufficiently high temperature, the first order phase
transition is described by the $O(3)$ symmetric sphaleron-type solutions. Thus the
Euclidean metric is
\begin{equation}\label{eme}
ds^{2}=\Bigl(1-\frac{2GM(r)}{r}\Bigr)e^{2\delta(r)}dt_E^{2}+\Bigl(1-
\frac{2GM(r)}{r}\Bigr)^{-1}dr^{2}+r^{2}(d\theta^{2}+\sin^{2}\theta
d\varphi^{2}),
\end{equation}
and the ansatz for the scalar field takes the form:
\begin{eqnarray}\label{sc}
\phi^{a}&=&\hat{\phi}^{a}(\theta,\varphi)\phi(r)\nonumber\\
&=&(\sin n\theta\cos n\varphi,\;\sin n\theta\sin n\varphi,\;\cos n\theta)
\,\phi(r),
\end{eqnarray}
where $n=0$ for the ordinary bubble solutions and $n=1$ for the
{\it monopole-bubble} solutions.

In addition to the well-known ordinary bubble solution ($n=0$) with the 
boundary conditions $d\phi/dr|_{r=0}=0$ and $\phi(r\rightarrow H^{-1})=0$, 
there exists another nontrivial bubble solution ($n=1$) which starts at
$\phi(r=0)=0$, reaches the maximum point $\phi(r=r_{{\rm turn}})=\phi_{{\rm turn}}$ and
goes back to $\phi(r\rightarrow H^{-1})=0$ (see Fig. 1).

\newpage


\setlength{\unitlength}{0.240900pt}
\ifx\plotpoint\undefined\newsavebox{\plotpoint}\fi
\sbox{\plotpoint}{\rule[-0.200pt]{0.400pt}{0.400pt}}%


\vspace{10mm}

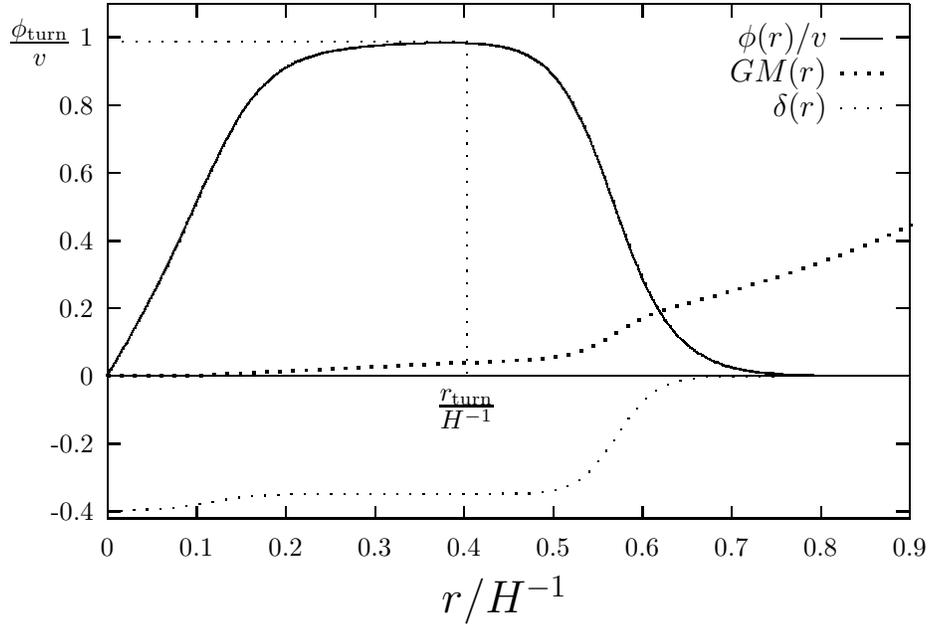
\noindent Figure 1. An $n=1$ bubble solution: The solid, dotted and dashed 
lines correspond to $\phi(r)$, $\delta(r)$ and $GM(r)$ when $\lambda=1$, 
$\alpha=0.1$ and $v/M_{Pl}=0.1$.

\vspace{10mm}


\setlength{\unitlength}{0.240900pt}
\ifx\plotpoint\undefined\newsavebox{\plotpoint}\fi
\sbox{\plotpoint}{\rule[-0.200pt]{0.400pt}{0.400pt}}%
%

\vspace{16mm}

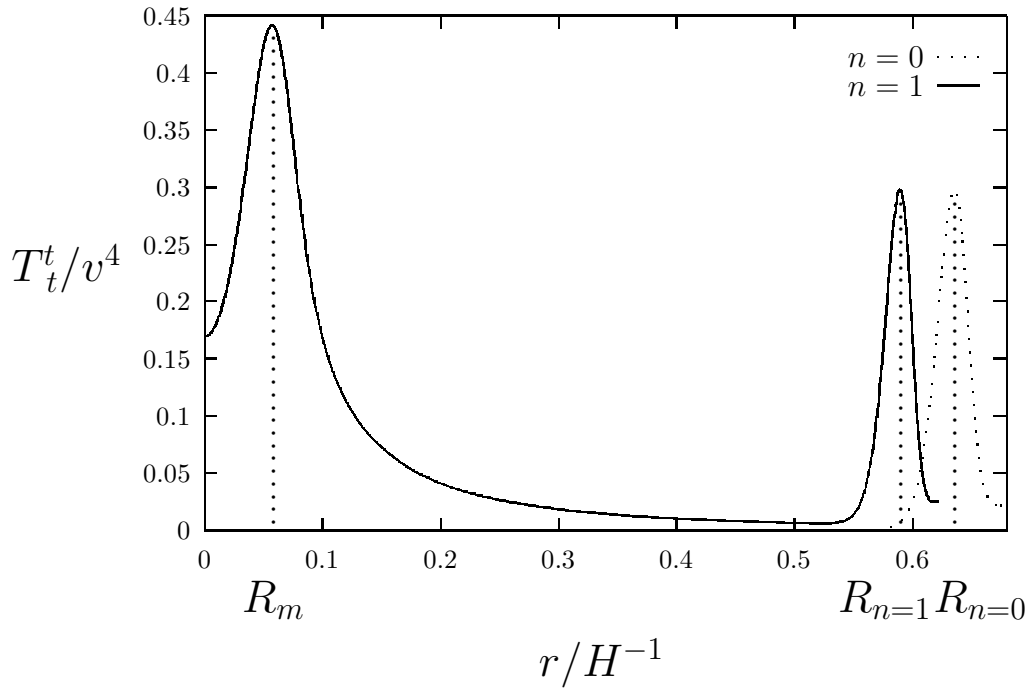
\noindent Figure 2. $T^{t}_{\;t}$ profiles for thin-wall bubble solutions: 
The dotted and solid lines correspond to an $n=0$ bubble and an $n=1$ bubble 
when $\lambda=1$, $\alpha=0.01$ and $v/M_{Pl}=0.1$.

\newpage

{}From the profile of $T^{t}_{\;t}$ after Wick rotation to Minkowski signature
in Fig. 2, we read the characteristics of the $n=1$ bubble in both flat and 
curved spacetimes:

\noindent (i) There are two bubble walls: One at $r/H^{-1}=R_{n=1}$ is the 
bubble wall which distinguishes the inside of the bubble at true vacuum from 
the outside of the bubble at false vacuum, and the other at $r/H^{-1}=R_{m}$ 
is that which divides the false vacuum core and the true vacuum surroundings.

\noindent (ii) The matter aggregate formed at the center of the $n=1$ bubble is nothing but the global monopole due to the nontrivial local mapping between the internal $O(3)$ symmetry and the spatial $O(3)$ symmetry. It is the reason why we call the $n=1$ bubble
the {\it monopole-bubble}.

\noindent (iii) There is a no-go theorem that says the scalar fields
described by the standard relativistic form of the Lagrangian do not support
any non-trivial static soliton solutions of finite energy. However, the global
monopole created inside the $n=1$ bubble is a finite energy configuration
since the long-range tail of the global monopole
\begin{equation}\label{eng}
T^{t}_{\;t}=\frac{1}{2}\Bigl(1-\frac{2GM}{r}\Bigr)\biggl(
\frac{d\phi}{dr}\biggr)^{2}+\delta_{n1}\frac{\phi^{2}}{r^{2}}+V
\stackrel{r\approx r_{{\rm turn}}}{\sim}\frac{\phi^{2}_{{\rm turn}}}{r^{2}},
\end{equation}
is tamed by the outer bubble wall at $r/H^{-1}=R_{n=1}$.

\vspace{5mm}

The gravitational effects on the shapes of the monopole-bubbles are:

\noindent (i) In terms of the radial coordinate $r$, $\phi(r)$ grows slowly
near the origin, and it reflects the repulsive nature of the gravity at the
global monopole core. The decrease of both $r_{{\rm turn}}$ and $\phi_{{\rm
turn}}$ can be  understood by the attractive nature of gravity in the true vacuum
region  between two walls.

\noindent (ii) In a flat spacetime the radius of an $n=1$ bubble is always
larger  than that of an $n=0$ bubble. This is also true for a thick wall
monopole-bubble in a curved spacetime. However, the size of a thin wall
monopole-bubble in a curved spacetime can be smaller than that of an $n=0$
bubble as shown in Fig. 2. Let us consider a junction condition for a thin wall
bubble \begin{equation}\label{junc}
\sqrt{\Bigl(\frac{dR}{d\tau}\Bigr)^{2}+1-\frac{8}{3}\pi
GV(0)R^{2}}-\sqrt{\Bigl(\frac{dR}{d\tau}\Bigr)^{2}+1-8\pi
Gv^{2} +\frac{2G|M_{{\rm turn}}|}{R}}=-4\pi G\sigma R, 
\end{equation}
where $R$, $\tau$, and $\sigma$ are  the circumference radius, the proper
time, and the surface energy density of the shell, respectively.
It tells us that the ratio of initial radii of the $n=0$ and $n=1$ bubbles can
be smaller than one for sufficiently large radius such that
\begin{equation}\label{R0}
\frac{R_{n=1}(0)}{R_{n=0}(0)}=\frac{1}{2}\left(\sqrt{1-8\pi Gv^{2}}+
\sqrt{1-8\pi Gv^{2}+\frac{4v^2}{R_{n=0}(0)\sigma}}\right)
\stackrel{R\rightarrow{\rm large}}{\longrightarrow}\sqrt{1-8\pi Gv^{2}}<1,
\end{equation}
where $R_{n=0}(0)\equiv3\sigma/(V(0)+6\pi G\sigma^2)$ is the initial radius of 
the $n=0$ bubble.

\vspace{5mm}

The spacetime structure inside and outside the monopole-bubble is summarized as
follows:

\noindent (i) The regions at the core of the global monopole
($r/H^{-1}<R_{m}$) and at the outside of the bubble ($r/H^{-1}>R_{n=1}$) are
de Sitter spacetimes described by the metric,
\begin{equation}\label{dS}
ds^{2}=-\Bigl(1-\frac{8}{3}\pi G V(0)r^{2}\Bigr)e^{2\delta_{0}}
dt^{2}+\Bigl(1-\frac{8}{3}\pi G V(0)r^{2}\Bigr)^{-1}dr^{2}
+r^{2}(d\theta^{2}+\sin^2\theta d\varphi^{2}).
\end{equation}

\noindent (ii) The true vacuum region between the inner and outer bubble
walls ($R_{m}<r/H^{-1}<R_{n=1}$) is a nearly flat spacetime with a deficit solid
angle $\Delta=8\pi Gv^{2}$. Specifically, after rescaling the variables $t$
and $r$, we have the metric near $r=r_{{\rm turn}}$,
\begin{eqnarray}\label{flat}
ds^{2}&=&-\Bigl(1-\frac{2GM_{{\rm turn}}}{1-8\pi G\phi^{2}_{{\rm turn}}}\frac{1}{r}\Bigr)
dt^{2}+\Bigl(1-\frac{2GM_{{\rm turn}}}{1-8\pi G\phi^{2}_{{\rm turn}}}\frac{1}{r}\Bigr)
dr^{2}\nonumber\\
&&+r^{2}(1-8\pi G\phi_{{\rm turn}}^{2})(d\theta^{2}+\sin^2\theta\,d\varphi^{2}),
\end{eqnarray}
where 
$$
GM_{{\rm turn}}=4\pi G\int^{r_{{\rm turn}}}_{0}
\!\!\!dr\,r^{2}\biggl\{\frac{1}{2}
\Bigl(1-\frac{2GM}{r}\Bigr)\biggl(\frac{d\phi}{dr}\biggr)^{2}\!
+\frac{\phi^{2}-\phi^{2}_{{\rm turn}}}{r^{2}}
+\bigl(V-V(\phi_{{\rm turn}})\bigr)\biggr\}.
$$
This flat nature of the metric can be understood in the Newtonian limit of the
Einstein equation: $\nabla^{2}\Phi=8\pi G(T^{t}_{\;t}-
T^{r}_{\;r}){\approx}0$ at $r\approx r_{{\rm turn}}$ \cite{BV,HL}.

\noindent (iii) Since the integral of the core region has the negative
contribution, {\it i.e.}, 
$M_{{\rm turn}}\approx -m_{H}=-\sqrt{4\lambda(3+2\alpha)}v<0$, the global
monopole inside the monopole-bubble does not form a black hole even at the
Planck scale \cite{HL}.

\noindent (iv) A currently open question is what is the structure of a
spacetime manifold which is formed when the deficit solid angle is equal to or
greater than $4\pi$.

\vspace{5mm}

When the system contains two distinct decay channels described by $n=0$ and 
$n=1$ bubbles, an interesting quantity is the ratio of two decay rates.
If the tunneling action for each bubble is larger than unity, the nucleation
rate for the $n^{{\rm th}}$ bubble takes the exponential form, {\it i.e.},
$\Gamma^{(n)}=A_{n}e^{-\frac{v}{T}B^{'}_{n}}$, where $B^{'}_{n}$ is the value of
the Euclidean action of the $n^{{\rm th}}$ bubble solution multiplied by $T/v$.
For the ratio of prefactors $A_{1}/A_{0}$, we assume that zero mode
contributions are dominant as has been done in a flat spacetime and then we have
\begin{equation}
\frac{\Gamma^{(1)}}{\Gamma^{(0)}}\sim\biggl(\frac{B^{'}_{1}}{B_{0}^{'}}
\biggr)^{\frac{6}{2}}\exp\Bigl[-\frac{v}{T}(B^{'}_{1}-B_{0}^{'})\Bigr].
\end{equation}
Note that the leading contribution of the action difference 
$(v/T)(B^{'}_{1}-B^{'}_{0})$ can be understood as the energy needed to generate
the global monopole at the center of monopole-bubble, which is approximately
proportional to the radius of the monopole-bubble, $B_{1}^{'}/B_{0}^{'}$ tends
to one in the thin-wall limit and a few in the thick-wall limit. Therefore the
substitution of the values of the action  gives several values of
$\Gamma^{(1)}/\Gamma^{(0)}$:

\begin{center}{
\begin{tabular}{|c||c|c|} \hline
$\alpha\;\backslash\;\frac{v}{T}$&1.0&2.0 \\ \hline\hline
0.3 (thick wall)&3.32$\times10^{-2}$&1.98$\times10^{-5}$\\ \hline
0.1&3.43$\times10^{-9}$&1.59$\times10^{-18}$\\ \hline
0.03 (thin wall)&5.23$\times10^{-13}$&1.47$\times10^{-25}$\\ \hline
\end{tabular}}
\end{center}

\begin{center}
{Table 1. Values of $\Gamma^{(1)}/\Gamma^{(0)}$ for $\lambda=1$ and
$v/M_{Pl}=0.1$.}
\end{center}

As expected, (i) monopole-bubbles are more likely to be nucleated at high
temperature and in the relatively thick wall case, and (ii) there may
exist some parameter region of a scalar potential where the monopole-bubbles
cannot be neglected, although $B_{1}^{'}$ is always larger than $B_{0}^{'}$.

\vskip 1cm
\noindent {\bf III. Evolution of Monopole-Bubbles}
\vskip 0.7cm

In order to pursue the evolution of bubbles, we solve the classical equations
of motion and take into account the effect due to the temperature change
when we prepare the initial conditions.  

\vspace{7mm}


\setlength{\unitlength}{0.240900pt}
\ifx\plotpoint\undefined\newsavebox{\plotpoint}\fi


\vspace{10mm}

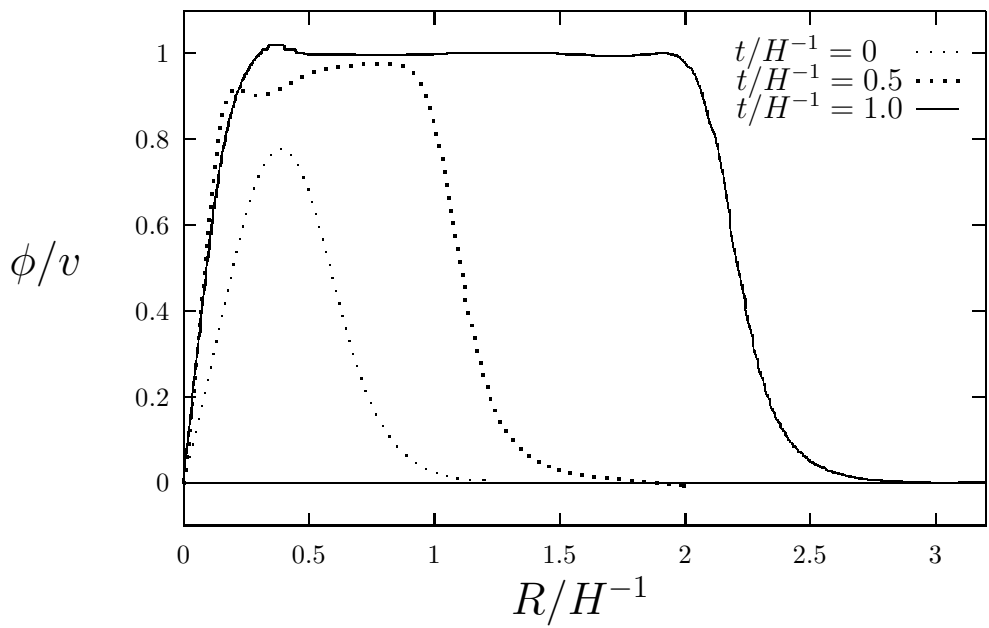
\noindent Figure 3. Profiles of an $n=1$ thick-wall bubbles for fixed 
$\lambda=1$, $\alpha=0.3$ and $v/M_{Pl}=0.1$ associated with the classical
evolution.

\vspace{5mm}

An example of our numerical results for the evolution of an $n=1$ bubble is
displayed in Fig. 3. Note that:

\noindent (i) When the size of the outer bubble wall is larger than that of the
static solution, {\it i.e.}, the critical size, the outer bubble wall of the
monopole-bubble starts to expand.

\noindent (ii) As the bubble grows, a thick-wall bubble becomes a thin-wall one.

\noindent (iii) If we trace the time-evolution of the position of $\phi=0.5v$
for the outer wall, its trajectory looks like a hyperbola, similar to that of
$n=0$ bubble wall. The terminal velocity of the wall is read from 
Eq.(\ref{junc}):
\begin{equation}\label{terv}
{\rm v_{terminal}}\approx 1-\frac{4\pi GV(0)}{3[R_{n=0}(0)]^2},
\end{equation}
which agrees with that of an $n=0$ bubble.

\noindent (iv) From the trajectory of the position of $\phi=0.5v$ for the inner
wall, we find that it oscillates and we can expect that this  oscillation will
be damped gradually. It implies that the global monopole inside an $n=1$ bubble
is stable against the perturbations of the scalar amplitude.

\vspace{5mm}

Finally, let us discuss the case where the scale of the symmetry breakdown $v$ 
approaches the Planck scale $M_{Pl}$. When the de Sitter horizon is smaller
than the radius of a monopole-bubble, one probable case is that the nucleated
bubble contains a super-horizon-sized monopole according to the viewpoint of
Ref.\cite{GW}. Once such a configuration is formed, one can expect defect
inflation at the monopole site \cite{Vil}.

\vskip 1cm  
We would like to thank R. Easther for reading the manuscript. Y.K.'s research was supported in part by the KOSEF (95-0702-04-01-3, Brain Pool Program, CTP of Seoul National University) and the Korean Ministry of Education (BSRI-95-2413). K.M. and N.S.'s research was supported in part by the Grant-in-aid for Scientific Research Fund of the Ministry of Education, Science and Culture (No.06302021, No.06640412, and No.07740226), and by the Waseda University Grant for Special Research Projects.

\end{document}